\documentclass[aps,prl,superscriptaddress,twocolumn]{revtex4}

\usepackage[pdftex]{graphicx}
\usepackage{dcolumn}
\usepackage{bm}
\usepackage{color}
\usepackage{amsmath}
\usepackage{centernot}
\usepackage{nicefrac}
\usepackage{amssymb} 

\newcommand{\ff}[1]{{\boldsymbol #1}}
\newcommand{\ca}[1]{{\cal #1}}
\newcommand{\bi}{\begin{itemize}}
\newcommand{\ei}{\end{itemize}}
\newcommand{\be}{\begin{equation}}
\newcommand{\ee}{\end{equation}}
\newcommand{\ba}{\begin{eqnarray}}
\newcommand{\ea}{\end{eqnarray}}

\newcommand{\titlepaper}{Interacting Chern Insulator in Infinite Spatial Dimensions}

\usepackage[pdftex,colorlinks=true,linkcolor=blue,citecolor=blue,filecolor=blue]{hyperref}

\begin{document} 
  
\title{\titlepaper}

\author{David Kr\"uger} 

\affiliation{I. Institute of Theoretical Physics, Department of Physics, University of Hamburg, Jungiusstra\ss{}e 9, 20355 Hamburg, Germany}

\author{Michael Potthoff}

\affiliation{I. Institute of Theoretical Physics, Department of Physics, University of Hamburg, Jungiusstra\ss{}e 9, 20355 Hamburg, Germany}

\affiliation{The Hamburg Centre for Ultrafast Imaging, Luruper Chaussee 149, 22761 Hamburg, Germany}

\begin{abstract}
We study a generic model of a Chern insulator supplemented by a Hubbard interaction in arbitrary even dimension $D$ and demonstrate that the model remains well-defined and nontrivial in the $D \to \infty$ limit. 
Dynamical mean-field theory is applicable and predicts a phase diagram with a {\em continuum} of topologically different phases separating a correlated Mott insulator from the trivial band insulator. 
We discuss various features, such as the elusive distinction between insulating and semi-metal states, which are unconventional already in the non-interacting case. 
Topological phases are characterized by a non-quantized Chern {\em density} replacing the Chern number as $D\to \infty$.
\end{abstract} 

\maketitle 

\paragraph{\color{blue} Introduction.}  
Strong electron correlations and topological classification are two major research frontiers of condensed-matter theory.
While much work has been done in providing prototypical examples of topologically nontrivial quantum matter \cite{HK10,QZ11,TKNN82,vK86} and in classifying
\cite{Zir96,AZ97,SRFL08,MF13,CTSR16} topological insulators, 
much less is known for correlated systems \cite{FK10,HLA11,YPFK14,Rac18}.
As correlated lattice-fermion models in $D=2$ and $D=3$ dimensions pose highly involved problems, many studies focus on one-dimensional systems with nontrivial topological properties \cite{TPB11,GS11,MENG12,GHF13,SMKS14,HSL+18,SK18}.

On the other hand, the opposite limit of infinite spatial dimensions has been recognized as extremely instructive for the pure electron-correlation problem and constitutive for the dynamical mean-field theory (DMFT) \cite{GKKR96}.
In the large class of mean-field approaches, DMFT has an exceptional standing, since it is internally consistent and nonperturbative, and since it becomes exact in the $D\to\infty$ limit \cite{MV89}.
While the limit comes with certain simplifications, such as the locality of the self-energy \cite{MH89a,GKKR96}, infinite-dimensional lattice-fermion models are far from being trivial. 
This is demonstrated by the DMFT paradigm of the Mott metal-insulator transition as a prime example \cite{Geb97}.
Furthermore, the fact that exact properties of strongly correlated systems are numerically accessible \cite{GKKR96,GML+11,LHGH14}, make correlated lattice-fermion models on $D = \infty$ lattices attractive points of orientation.

With the present study we pose the question whether the same limit is also helpful for the understanding of topological properties of strongly interacting electron systems. 
Our answer is affirmative.
Assuming locality of the self-energy, 
previous DMFT studies of correlated topological insulators have addressed two-dimensional systems, such as the 
Haldane model \cite{VSL+16}, 
Hofstadter's butterfly \cite{MRR19}, or the 
BHZ model \cite{WDX12,ABC+15}, all supplemented by interaction terms, 
or real three-dimensional systems, such as SmB$_{6}$ \cite{TH19}, combining the DMFT with ab initio band theory.
A DMFT study of an interacting, topologically nontrivial model on a $D = \infty$ lattice is still missing.

Here, we consider multi-orbital Hubbard models on a $D$-dimensional hypercubic lattice for arbitrary but even $D$, whose low-energy non-interacting band structures reduce to massive Dirac theories and belong to class A  of Chern insulators with $\mathbb{Z}$ topological invariants.
We demonstrate that, with the proper scaling of the hopping, the $D\to \infty$ limit leads to a well-defined model with nontrivial interplay between kinetic and interaction terms, hosting topologically nontrivial phases, and is accessible to a numerical solution by DMFT for arbitrary Hubbard interaction $U$ and mass  parameter $m$. 
The $m$-$U$ phase diagram contains the trivial band and the correlated Mott insulator, separated by a {\em continuum} of interacting and topologically different Chern insulators. 
The latter are characterized by a properly defined Chern {\em density}, which replaces the Chern number as a topological invariant.
We argue that for $D\to \infty$ already the $U=0$ model has highly unconventional topological properties as the sign of the Chern number as well as a band closure are concepts becoming ill-defined in the limit $D\to \infty$.

\paragraph{\color{blue} Hamiltonian.}  
We study an extension of a family of $D$-dimensional tight-binding models for even $D$ to spinful fermions with local Coulomb interaction as described by the Hamiltonian $H = H_{0}+ H_{1}$. 
Here $H_{1} = (U/2) \sum_{i\alpha\sigma} n_{i\alpha\sigma} n_{i \alpha -\sigma}$ is an on-site and intra-orbital Hubbard term, where $i=1,...,L$ labels the sites of a $D$-dimensional hypercubic lattice with periodic boundaries, $\sigma = \uparrow,\downarrow$ is the spin projection, and $\alpha = 1, ..., M$ is an orbital index.
The corresponding annihilator is $c_{i \alpha \sigma}$, and $n_{i\alpha\sigma} \equiv c^{\dagger }_{i \alpha \sigma}c_{i \alpha \sigma}$.
After Fourier transformation to $k$-space, $c_{i\alpha\sigma} = L^{-1/2} \sum_{k} e^{ ikR_{i}} c_{k\alpha\sigma}$, the tight-binding part reads 
$H_{0} = \sum_{k \alpha \beta \sigma} \epsilon_{\alpha\beta}(k) c_{k\alpha\sigma}^{\dagger}c_{k\beta\sigma}$, where $k=(k_{1},...,k_{D})$ with $-\pi < k_{r} \le \pi$, and where $\epsilon_{\alpha\beta}(k)$ are the elements of the $M \times M$ hopping matrix in k-space:
\be
  \ff \epsilon(k)
  =
  \left(
  m + t \sum_{r=1}^{D} \cos k_{r}
  \right)
  \ff \gamma_{D}^{(0)}
  +
  t 
  \sum_{r=1}^{D} \sin k_{r} 
  \ff \gamma_{D}^{(r)} 
   \: ,
\label{hk}
\ee
depending on the hopping parameter $t$ and on a parameter $m$ controlling the mass term.
Here, $\gamma_{D}^{(1)}, ..., \gamma_{D}^{(D)}$ are the generators of the complex Clifford algebra $\mathbb{C}l_{D}$, and $\gamma_{D}^{(0)} = (-i)^{D/2} \gamma_{D}^{(1)} \cdots \gamma_{D}^{(D)}$ is the chiral element.
They satisfy the Clifford anticommutation relations 
$\{ \ff \gamma_{D}^{(\mu)} , \ff \gamma_{D}^{(\nu)} \} = 2 \delta^{(\mu\nu)}$ for $\mu, \nu= 0,1,...,D$.
Close to the critical points $k_{\rm c}$ in the first Brillouin zone (BZ), see below, the low-energy effective theory is given by a linear Dirac model with $k$-independent mass term.
Such free Dirac models are extensively analyzed and topologically classified for different mass terms and for arbitrary $D$, see e.g.\ Ref.\ \cite{PSB16}.
The model (\ref{hk}) belongs to symmetry class A in the Altland-Zirnbauer (AZ) scheme \cite{AZ97}.

We note that $\mathbb{C}l_{D+2} \cong \text{Mat}(2,\mathbb{C}) \otimes \mathbb{C}l_{D}$ and that there is, for even $D$, a unique irreducible $M=2^{D/2}$-dimensional matrix representation of $\mathbb{C}l_{D}$ \cite{Lan02,Tra05,BKU06}. 
The according $\gamma$-matrices can be constructed recursively: 
$\mathbb{C}l_{0}$ is spanned by $1 \in \mathbb{C}$.
The first nontrivial dimension is $D=2$, and hence $M=2$.
$\mathbb{C}l_{2}$ is generated by the Pauli matrices $\ff \gamma_{2}^{(1)} = \ff \tau_{x}$ and $\ff \gamma_{2}^{(2)} = \ff \tau_{y}$, and together with the unity $\ff 1$ and the chiral element $\ff \gamma_{2}^{(0)} = -i \ff \tau_{x} \ff \tau_{y} = \ff \tau_{z}$, they span $\mathbb{C}l_{2}$. 
The corresponding generalized lattice Dirac model, Eq.\ (\ref{hk}), with $\ff \epsilon(k) = d(k) \cdot \ff \tau$ and $d(k) = (t \sin k_{x},t \sin k_{y}, m + t\cos k_{x} + t\cos k_{y})$ is just the model proposed by Qi, Wu and Zhang \cite{QWZ06,AOP16}.
For arbitrary even $D$ the general recursive prescription for the Hermitian and traceless generators is \cite{PSB16}:
\ba
  \ff \gamma_{D+2}^{(r)} &=& \ff \tau_{x} \otimes \ff \gamma_{D}^{(r)} \: , \; \mbox{for} \: r=1,...,D \nonumber \\
  \ff \gamma_{D+2}^{(D+1)} &=& \ff \tau_{x} \otimes \ff \gamma_{D}^{(0)} \: , \:
  \ff \gamma_{D+2}^{(D+2)} = \ff \tau_{y} \otimes \ff 1 \: .
\label{eq:gamma}
\ea
The chiral element is $\ff \gamma_{D+2}^{(0)} = \ff \tau_{z} \otimes \ff 1$, where $\ff 1$ denotes the $2^{D/2}$-dimensional unity. 
Explicitly, $\ff \gamma^{(0)} = \mbox{diag}(+1,+1, ..., -1,-1,...)$, such that $m$ is the strength of a staggered on-site potential in Eq.\ (\ref{hk}).
Accordingly, the orbitals $\alpha$ can be divided into two classes, A orbitals with $\gamma^{(0)}_{\alpha\alpha} \equiv z_{\alpha} = +1$ ($\alpha=1,...,M/2$) and B orbitals  
$\gamma^{(0)}_{\alpha\alpha} \equiv z_{\alpha} = -1$ ($\alpha=(M/2) + 1,...,M$).
We see that the number of orbitals scales exponentially with $D$.
Eqs.\ (\ref{hk}) and (\ref{eq:gamma}) imply that along a spatial direction $r$, each site-orbital $(i,\alpha)$ couples to a single orbital $\alpha'$ at the two nearest-neighbor positions $i'$, and thus the connectivity of $(i,\alpha)$ is $2D$.

\paragraph{\color{blue} Noninteracting case.}  
The $U=0$ band structure is easily obtained by squaring $\ff \epsilon(k)$, using properties of the $\ff \gamma$ matrices, and noting that $\mbox{tr}\, \ff \epsilon(k) = 0$.
Apart from the spin degeneracy, this yields two $M/2$-fold degenerate bands:
$\epsilon_{\pm}(k) = \pm [ t^{2} \sum_{r} \sin^{2}k_{r} + (m+ t \sum_{r} \cos k_{r})^{2} ]^{1/2}$.
The high-energy band edges are given by $\epsilon_{\rm max, min} = \pm (|m|+Dt)$ and are taken for $k_{r}=0$ (if $m\ge 0$) and $k_{r}=\pi$ ($m\le 0$) for all $r$.
Due to the point-group symmetries, band closures are found at the high-symmetry points (HSPs) $k_{\rm c} = k_{n_{0}} = (0,...,0,\pi, ..., \pi)$ in the BZ, and for $D \choose n_{0}$ inequivalent permutations of the components, where $n_{0}$ counts the number of vanishing entries $k_{r}$.
For a band closure the condition $m=(D-2n_{0})t$ must be met.
This corresponds to the vanishing of the mass term in the Dirac Hamiltonian
$\ff \epsilon(k) = [m+(2n_{0}-D)t] \ff \gamma_{D}^{(0)} + t \sum_{r} (k_{r} - k_{n_{0},r}) \ff \gamma^{(r)}_{D}$, obtained by linearization of $\ff \epsilon(k)$ around $k_{n_{0}}$.

%%%%%%%%%%%%%%%%%%%%%%%%%%%%%%%%%%%%%%%%%%%%%%%%%%%
\begin{figure*}[t]
\includegraphics[width=1.9\columnwidth]{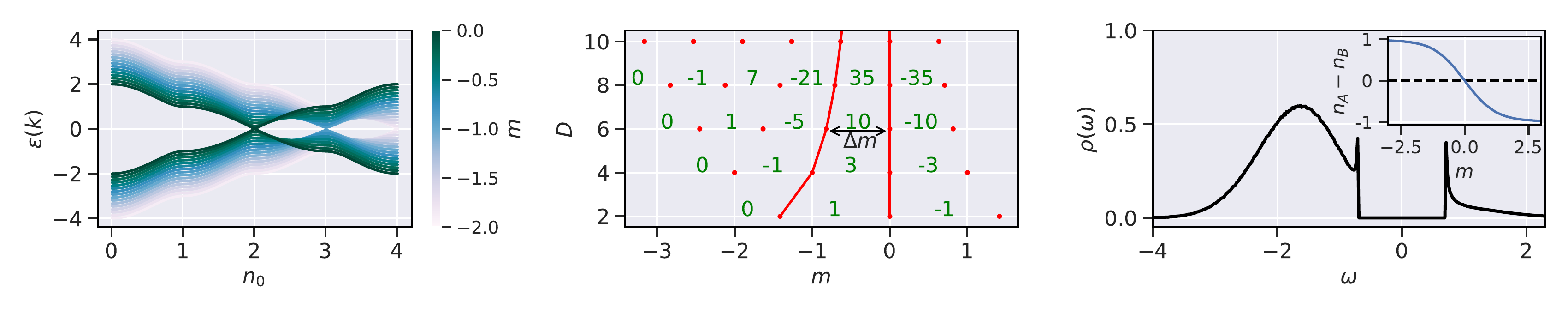}
\caption{
{\em Left:} Band structure $\epsilon(k)=\epsilon_{\pm}(k)$ of the $D=4$ model along straight shortest lines in the BZ connecting HSPs characterized by $n_{0}$. Results for different $m$, see color code. 
{\em Middle:} Different topological phases with Chern numbers $C_{D}(n_{0})$ (green), separated by critical $m$-values (red dots) for different $D$.
{\em Right:} $U=0$ DOS on the A-orbitals at $m=-1.5$ for $D=\infty$.
Inset: orbital polarization as function of $m$ for $D=\infty$.
Nearest-neighbor hopping: $t = t^{\ast} / \sqrt{D}$, $t^{\ast}=1$ sets the energy scale.
}
\label{band}
\end{figure*}
%%%%%%%%%%%%%%%%%%%%%%%%%%%%%%%%%%%%%%%%%%%%%%%%%%%

\paragraph{\color{blue} Infinite dimensions.}  
It is instructive to compute the low-order moments
$M_{\alpha}^{(n)} = \int d\omega \,\rho^{n}_{\alpha}(\omega)$
of the local partial density of states (DOS) of the orbital $\alpha$.
We have the trivial normalization condition $M_{\alpha}^{(0)}=1$, the barycenter 
$M_{\alpha}^{(1)}=m \gamma^{(0)}_{\alpha\alpha} = \pm m$, and the $\alpha$-independent second moment $M_{\alpha}^{(2)}=t^{2}D+m^{2}$.
The variance of the DOS is given by the second central moment $M_{\alpha}^{(2)} - (M_{\alpha}^{(1)})^{2} = t^{2} D$. 
Hence, a proper $D\to \infty$ limit with a balance between $H_{0}$ and $H_{1}$ is obtained if the standard \cite{MV89,MH89b} scaling $t=t^{\ast} / \sqrt{D}$ with lattice dimension $D$ is employed.
This will be assumed here as well.
Furthermore, we fix the energy scale by setting $t^{\ast}=1$, i.e., the variance of the DOS is unity, while the locations of the band edges diverge $\epsilon_{\rm max, min} = \pm (|m|+\sqrt{D}t^{\ast}) \mapsto \pm \infty$.
The mass parameter $m$ must not be scaled in the $D\to \infty$ limit to maintain a nontrivial model.
This implies a $D$-independent band center of gravity $\pm m$.

\paragraph{\color{blue} Topology for $D\to\infty$.}  
We approach the $D \to \infty$ limit via even-$D$ models of Chern insulators and stay in the AZ class A.
For any finite even $D$, upon varying $m$, one passes band closures and related topological phase transitions, located at $m=\sqrt{D}(1-2n_{0}/D) t^{\ast}$ for $n_{0}=0,...,D$. 
Fig.\ \ref{band} (left) gives an example for $D=4$. 
The topological phase for an $m$ with $D-2n_{0}-2 < m \sqrt{D}/t^{\ast} < D - 2 n_{0}$ (with $n_{0}=0,...,D-1$) can be characterized by the $(D/2)$-th Chern number \cite{PSB16,GJK93,QHZ08}:
\be
  C_{D}(n_{0}) = (-1)^{n_{0} +\frac D2} {D-1 \choose n_{0}} \: , 
\label{ch}  
\ee
see Fig.\ \ref{band} (middle) for an overview.
The equation can be interpreted by referring to the bulk-boundary correspondence \cite{SRFL08}.
Namely, the $(D-1)$-dimensional surface characterized by Miller indices $(100 \cdots 0)$ hosts topologically protected surface states, and their dispersion has Weyl nodes at the surface projections $k_{{\rm c,} \|}$ of the bulk HSPs $k_{\rm c}$ for given $n_{0}$. 
The binomial factor in Eq.\ (\ref{ch}) counts the number of equivalent nodal $k_{{\rm c,} \|}$-points in the $(D-1)$-dimensional surface Brillouin zone. 
All Weyl points have the same chirality given by the sign factor \cite{PSB16}.

Importantly, the distance between two transitions $\Delta m = 2 t^{\ast} / \sqrt{D}$ shrinks to zero for $D\to \infty$, i.e., the set of critical $m$'s becomes dense in any finite $m$-interval.
Hence, for high $D$ the system is arbitrarily close to criticality for {\em any} $m$.
We note that, mathematically, the definition of a critical point in the BZ becomes elusive for $D\to \infty$, since $\epsilon_{\pm}(k) = \epsilon_{\pm}(k')$ if $\| k - k' \| = 0$, where we have defined $\| k \|^{2} \equiv \lim_{D\to \infty} D^{-1} \sum_{r=1}^{D} k_{r}^{2}$. 
It is easy to see that $\| \cdot \|$ is a semi-norm, i.e.\ $\| k \| = 0 \not \Rightarrow k=0$, such that the concept of a band closure at isolated points in $k$-space breaks down.
However, we still have $\epsilon_{\pm}(k) = 0$ at $k=k_{\rm c}(m)$ for any $m$.
Furthermore, the number $D \choose n_{0}$ of equivalent critical HSPs at a given critical $m$ and the total number $2^{D}$ of HSPs in the BZ diverge, but their ratio approaches a constant when $D\to \infty$.

A second important observation directly follows from Eq.\ (\ref{ch}): 
When $D\to \infty$, only the {\em modulus} of the Chern number, and only after proper {\em normalization}, has a well-defined limit.
Noting that $\sum_{n_{0}=0}^{D-1} C_{D}(n_{0}) = 2^{D-1}$, we thus introduce a Chern {\em density} as $c(n_{0}) = \lim_{D\to\infty} |C_{D}(n_{0)}|/2^{D-1}$.
Since $\Delta m \mapsto 0$, we can use $n_{0}=(D - m\sqrt{D}/t^{\ast})/2$ and $dm \equiv \frac{2t^{\ast}}{\sqrt{D}}$ to express the Chern density as a function of $m$.
With this, and using the Moivre-Laplace theorem, we find:
\be
   c(n_{0})
   = 
   \lim_{D\to\infty} \sqrt{\frac{2}{\pi D}} e^{-2\frac{(\frac D2 - n_{0})^{2}}{D}}
   =
  c(m) dm
\ee 
with a normalized Chern density of unit variance:
\be
c(m) =\frac{1}{t^{\ast}\sqrt{2\pi}} e^{-\frac{1}{2} \frac{m^{2}}{t^{\ast 2}}} \: .
\label{eq:gauss}
\ee
This is a central result, as it shows that not only dynamic correlation effects but also nontrivial topological properties survive the $D\to \infty$ limit when using the standard scaling of the hopping. 

From the bulk-boundary correspondence \cite{SRFL08,EG11,PSB16} at any finite $D$, we can infer that $c(m)dm$ is the {\rm ratio} between the number of topologically protected surface states and the total number of HSPs in the BZ.
Upon variation of $m \mapsto m+dm$, a ratio of $\pm 2 c(m)dm$ bulk states (per total number of HSPs) traverse the gap at the HSPs corresponding to $m$.
The Chern density is insensitive to the sign though.

\paragraph{\color{blue} Density of states.}  
Turning to the correlation side of the problem, the relevant quantity for the DMFT is the $U=0$-DOS $\rho_{\alpha}(\omega)=-(1/\pi L)\mbox{Im} \sum_{k} G^{(0)}_{\alpha\alpha}(k,\omega+i0^{+})$ of orbital $\alpha$.
This can be computed efficiently using the quasi-Monte Carlo technique of Refs.\ \cite{KN16,EP16} to carry out the $k$-summation. 
Thanks to the Clifford algebra, the inversion of the $M \times M$ hopping matrix required to get the noninteracting Green's function matrix $\ff G_{k}^{(0)}(\omega)=1/(\omega - \epsilon(\ff k))$ can be done analytically, see section A of the Supplemental Material (SM) \cite{SM}.
We also derive an analytical expression for the DOS in the $D\to \infty$ limit (SM, Sec.\ B \cite{SM}).
For any $D$, we have $\rho_{A}(-\omega) = \rho_{B}(\omega)$, and for $m \mapsto - m$, the DOS transforms as $\rho_{\alpha}(\omega) \mapsto \rho_{\alpha}(-\omega)$.
The $D=\infty$ DOS is shown in Fig.\ \ref{band} (right).

Another important point is that the $D=\infty$ DOS is fully gapped {\em for all} $m$.
Furthermore, the gap $\Delta=\sqrt{2} t^{\ast}$ is $m$-independent.
This should be contrasted with the DOS at any finite $D$, which behaves at low frequencies and at a critical $m$ as $\rho_{\alpha}(\omega) \propto |\omega|^{D-1}$, as it is characteristic for a Dirac-cone structure  (SM, Secs.\ C \cite{SM}). 
The band states near a band closure in $k$-space at a critical $k_{\rm c}$ and all equivalent points (including $k$-points with $\|k-k_{\rm c}\|$=0) do no longer contribute a finite DOS near $\omega=0$. 
Hence, there is no meaningful distinction between insulator and semi-metal states in the $D\to \infty$ limit.

The relevant range of the mass parameter to get nontrivial correlation effects in high $D$ is of order $m = \pm \ca O(t^{\ast})$.
This is demonstrated with the inset of Fig.\ \ref{band} (right) showing the orbital polarization $p=(n_{A} - n_{B})/2$ of the half-filled noninteracting system, $(n_{A} + n_{B})/2=1$, as a function of $m$ (where $n_{\alpha} \equiv L^{-1} \sum_{k\sigma} \langle c_{k\alpha\sigma}^{\dagger}c_{k\alpha\sigma} \rangle$).
On the scale $m = \pm \ca O(t^{\ast})$, $p$ quickly approaches almost full saturation with empty or doubly occupied A (or B) orbitals, i.e., a state where the Hubbard interaction is static and correlation effects are absent.

\paragraph{\color{blue} DMFT.}  
The exact solution of the interacting model in the $D\to \infty$ limit is provided by the DMFT. 
Particularly, the $m$-$U$ phase diagram of the model is interesting as it expresses the generic interplay of topological properties {\em and} correlations in an exactly solvable and non-perturbative case.
To cover the entire relevant parameter space, we employ a simplified DMFT scheme, where the interacting lattice model is self-consistently mapped onto a two-site single-impurity Anderson model (SIAM) \cite{Pot01a}.
A slight generalization is necessary to account for the A-B orbital structure.
This generalized two-site DMFT (see SM, Secs.\ D and E for details \cite{SM}) simultaneously focusses on the low- and on the high-frequency limit of the DMFT self-consistency condition and qualitatively captures the Mott-transition physics \cite{Pot01a,BP00,SPN03}.

At finite $U$ the $D$-th Chern number can be expressed in terms of the interacting single-particle Green's function \cite{IM86,WQZ10,WZ12}.
Here, for $D\to \infty$, the locality of the self-energy allows us to apply the concept of the topological Hamiltonian \cite{WZ12} (see also Ref.\ \cite{HWML16}) and to compute $c(m)$ from the noninteracting part but with $\ff \epsilon(k) \mapsto \ff \epsilon(k) - \mu \ff 1 + \ff \Sigma(\omega=0)$ and where the chemical potential $\mu = U/2$. 
Since $\ff \Sigma(\omega)$ is diagonal in orbital space (see SM, Sec.\ D \cite{SM}), this merely amounts to a renormalization of the chemical potential, $\mu \mapsto \mu + \Sigma_{+}(\omega=0)$, and the mass parameter, $m \mapsto m + \Sigma_{-}(\omega=0)$, where
$\Sigma_{\pm}(\omega)=(\Sigma_{A}(\omega) \pm \Sigma_{B}(\omega))/2$.
For finite $D$ we have successfully tested our results case by case against the predictions of the pole-expansion technique \cite{SHK06,WDX12,TH19}, which applies if the self-energy is given in its discrete Lehmann representation \cite{GP15}.

%%%%%%%%%%%%%%%%%%%%%%%%%%%%%%%%%%%%%%%%%%%%%%%%%%%
\begin{figure}[t]
\includegraphics[width=0.8\columnwidth]{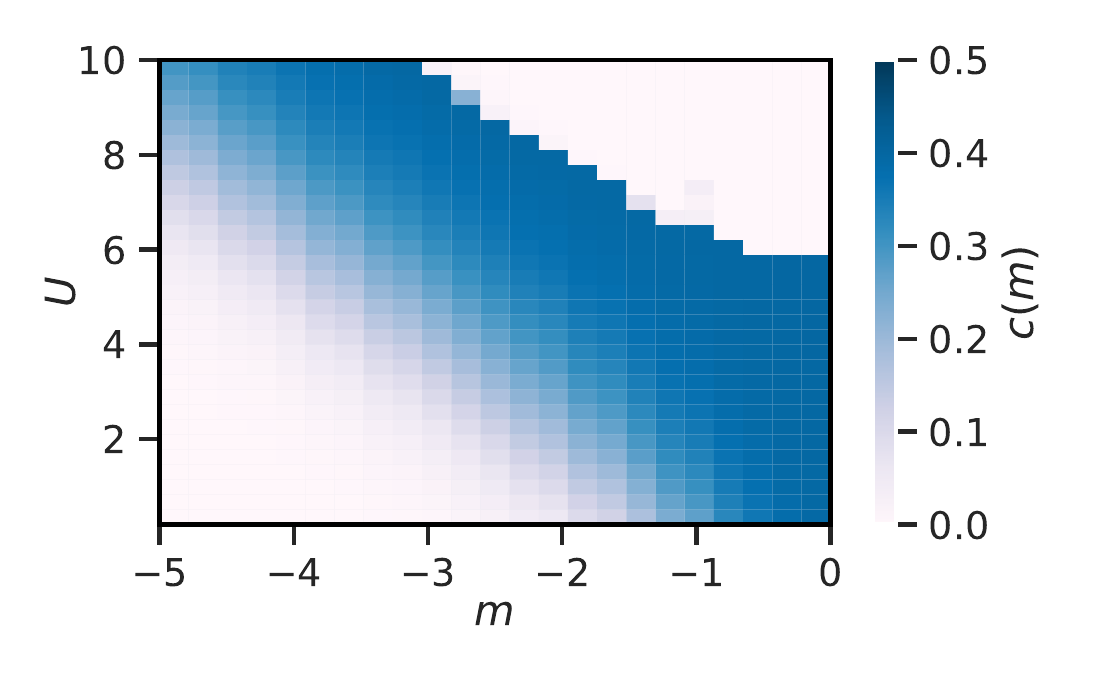}
\caption{
$m$-$U$ phase diagram of the $D\to \infty$ model. 
The color codes the Chern density.
}
\label{pd}
\end{figure}
%%%%%%%%%%%%%%%%%%%%%%%%%%%%%%%%%%%%%%%%%%%%%%%%%%%

\paragraph{\color{blue} Phase diagram.}  
We have performed DMFT calculations, restricted to spin-symmetric states in a large range of parameters $m$ and $U$. 
The resulting Chern density $c(m,U)$ is shown in Fig.\ \ref{pd}. 
As the phase diagram is invariant under a sign change $m\mapsto -m$, only negative $m$-values are displayed.
At $U=0$ and as a function of $m$, the Chern density is a Gaussian, see Eq.\ (\ref{eq:gauss}), and the system smoothly evolves from a conventional band insulator, with $c(m,0) \to 0$ in the limit $m \to -\infty$, to a Chern insulator / semimetal with a maximum $c(m,0)=1/\sqrt{2\pi}$ at the symmetric point $m=0$.

With increasing $U$ at $m=0$, the Chern density $c(0,U)$ stays at its maximum until at $U = U_{\rm c} = 6 t^{\ast}$, the system undergoes a correlation-driven transition to a topologically trivial Mott phase with $c=0$.
With a refined DMFT scheme only a slightly lower $U_{\rm c}$ is expected \cite{Pot01a}.
Approaching $U_{\rm c}$ either from above or from below, the transition is characterized by a continuously vanishing renormalization factor $z \mapsto 0$, where $z \equiv 1/(1-\partial \Sigma_{\alpha}(\omega=0)/\partial \omega)$ is independent of the orbital type $\alpha$.
$z$ plays the role of a band-gap renormalization \cite{SKWK09}.

The Mott phase extends to $m<0$ and is bounded for all $m$ by a line of critical interactions $U_{\rm c}(m)$. 
For $m\to -\infty$ we observe that $U_{\rm c}(m)$ linearly increases with $|m|$. 
This is explained by the fact that the system becomes fully orbital-polarized.
Hence, the self-energy becomes static and approaches constants $\Sigma_{A} \to U$, $\Sigma_{B} \to 0$, such that the renormalization of $m$ is trivial: $m \to m + \Sigma_{-}(\omega=0) \to m + U/2$.
As a consequence, the band insulator with $c=0$ cannot be smoothly connected to the Mott insulator with $c=0$ without passing topologically nontrivial states with $c>0$.

The whole phase diagram can be understood as the $D\to \infty$ limit of $m$-$U$ phase diagrams at finite $D$, see SM, Sec.\ F \cite{SM}.
With increasing $D$, the number of topologically nontrivial phases $C_{D}(m)\ne 0$ increases and become ever narrower regions in the $m$-$U$ plane, until they shrink to one-dimensional lines (of constant color in Fig.\ \ref{pd}) given by $c(m,U)=\mbox{const}$. 
This implies that, in the limit $D\to \infty$, systems on these iso-Chern curves are topologically equivalent, while on paths crossing iso-Cherns one passes through a continuum of topologically different phases. 

\paragraph{\color{blue} Conclusions and outlook.}  
DMFT is nowadays mostly employed as an approximate approach to strongly correlated lattice-fermion models in low dimensions. 
The fact that DMFT becomes exact in the $D\to \infty$, however, is a central aspect of the approach, as it ensures its internal consistency in the entire parameter space spanned by hopping, interaction, filling, orbital hybridization, and more.
It is thus important to demonstrate the very existence of an infinite-dimensional interacting lattice model with nontrivial topological properties that is in fact exactly solved by the DMFT.
The generic model of an interacting Chern insulator studied here is the first example of this kind.

Our approach has shown that with the conventional scaling of the hopping parameter, a nontrivial interplay between strong local correlations and topological properties is retained in the $D\to \infty$ limit and thus paves the way for further {\em generic} studies of this and of other models, including models in different AZ classes.
Such studies offer the unique possibility to exactly access intertwined correlation and topological effects in a nonperturbative regime and support approximate DMFT studies of low-$D$ cases.
%as well as complement numerically exact, e.g., matrix-product-state approaches of interacting $D=1$ systems.
They furthermore disentangle the pure and generic (dynamical) mean-field content of the theory from the additional realistic features of the DMFT when applied to low-$D$ models with specific lattice and orbital structure.
Clearly, also controlled ``expansions'' around the $D\to \infty$ limit, using cluster or diagrammatic schemes, profit from a well-defined and nontrivial starting point, e.g., for benchmarking.

%Another interesting route to proceed is to analyze the topologically protected and interacting surface states hosted by the model studied here. 
%This is worthwhile as we have seen that the information on the sign of the Chern number and thus on their chirality is lost in $D\to \infty$ limit. 
%Real-space DMFT for infinite-dimensional slab geometries can be employed here.

The question what is generic and what is specific in the context of interacting mean-field theory can also be posed with respect to the topological invariant itself.
For the model studied here, there is a {\em continuum} of topologically different phases, characterized by a Chern {\em density}, which is a {\em smooth}, non-quantized function of $m$ and $U$, except at the Mott transition.
Importantly, one can further elaborate on these ideas already in $U=0$ limit. 
While this provokes the question whether analogs can be found in finite-$D$ models, these features are interesting in themselves and one may even speculate about a possibly different topological classification in the $D\to \infty$ limit.

%There is a point that is specific to the $D\to \infty$ limit: 
%For any $m$, the DOS stays fully gapped while {\em at the same time} there is a band closure at a critical point $k_{\rm c}=k_{\rm c}(m)$ in the BZ, which, in the high-$D$ limit, should actually be seen as a representative of an equivalence class of $k$-points with zero mutual distance, derived from the semi-norm $\| \cdot \|$.
%Hence, while the model has robust topological properties, the distinction between insulating and semi-metal states becomes elusive. 

%One lesson we learnt in the early days of DMFT is that there is no direct feedback of nonlocal correlations on the self-consistent construction of effective impurity model.
%Here, with respect to topological properties, there is a similar lesson: 
%There is no direct feedback of the low-energy and local-in-$k$-space electronic structure and thus of  topological transitions on the DMFT self-consistency, and only the (gapped) DOS is relevant for the construction of effective impurity model generating the self-energy.

At the fundamental level of topological classification, there is obviously a plethora of open questions, including the robustness against including interactions \cite{FK11}, relevance of periodicity in the spatial dimension for interacting systems \cite{GJF19}, etc. 
Furthermore, we note that there are other routes to topological phases as well:
Nontrivial topological states could be generated by starting from a topologically trivial model in the $D\to \infty$ limit, either at some finite $D$ or via extensions of DMFT. 
Exact statements or even the exact construction of entire phase diagrams and of excitations spectra, however, are probably difficult to achieve beyond the dynamical mean-field concept but highly desirable. 

%--------------------------------------------------------------------------------------------------------------
\paragraph{\color{blue} Acknowledgments.}  
This work was supported by the Deutsche Forschungsgemeinschaft (DFG) through the Cluster of Excellence ``Advanced Imaging of Matter'' - EXC 2056 - project ID 390715994, and by the DFG 
Sonderforschungsbereich 925 ``Light-induced dynamics and control of correlated quantum systems''
(project B5).

%\end{document}

%%%%%%%%%%%%%%%%%%%%%%%%%%%%
%%%%%%%%%%%%%%%%%%%%%%%%%%%%
%%%%%%%%%%%%%%%%%%%%%%%%%%%%
%%%%%%%%%%%%%%%%%%%%%%%%%%%%
%%%%%%%%%%%%%%%%%%%%%%%%%%%%
%%%%%%%%%%%%%%%%%%%%%%%%%%%%

%Suppl. Mat.:

\begin{widetext}
\newpage
\mbox{}
\setcounter{page}{1}
%%%%%%%%%%%%%%%%%%%%%%%%%%%%

\begin{center}
{\large \bfseries 
\titlepaper

\mbox{}\\

--- Supplemental Material ---
}

\mbox{}\\

{
David Kr\"uger$^{1}$
and
Michael Potthoff$^{1,2}$
}

\mbox{}\\[-2mm]

{
\small \it 
$^{1}$I. Institute of Theoretical Physics, Department of Physics, 

University of Hamburg, Jungiusstra\ss{}e 9, 20355 Hamburg, Germany

$^{2}$The Hamburg Centre for Ultrafast Imaging, Luruper Chaussee 149, 22761 Hamburg, Germany
}

\end{center}

\end{widetext}

%%%%%%%%%%%%%%%%%%%%%%%%%%%%
\paragraph{\color{blue} 
Section A: Band structure and density of states.}

Here, we discuss some properties of the non-interacting part of the Hamiltonian. 
A Julia script for the computation of the density of states as function of $m$ and $D$ is available from the authors upon request.

With $d_{0}(k) \equiv  m + t \sum_{r=1}^{D} \cos k_{r}$ and $d_{r}(k) \equiv t \sin k_{r}$ for $r=1,...,D$ the $M\times M$ dispersion matrix $\ff \epsilon(k)$ (with $M=2^{D/2}$) in Eq.\ (\ref{hk}) can be written as
$\ff \epsilon(k) = d_{0}(k) \ff \gamma_{D}^{(0)} + \sum_{r=1}^{D} \sin k_{r} \ff \gamma_{D}^{(r)}$. 
Using the Clifford anticommutation relations 
$\{ \ff \gamma_{D}^{(\mu)} , \ff \gamma_{D}^{(\nu)} \} = 2 \delta^{(\mu\nu)}$, one finds
$\ff \epsilon(k)^{2} = (d_{0}(k)^{2} + \sum_{r} d_{r}(k)^{2}) \ff 1$, where $\ff 1$ is the $M$-dimensional unity.
The $\gamma$-matrices and thus $\ff \epsilon(k)$ are traceless. 
Hence, disregarding the spin degree of freedom, there are two $M/2$-fold degenerate bands with dispersions given by
$\epsilon_{\pm}(k) = \pm (d_{0}(k)^{2} + \sum_{r} d_{r}(k)^{2})^{1/2}$.

The gap $\Delta = \mbox{min}_{k} (\epsilon_{+}(k) - \epsilon_{-}(k))$ is determined by the conditions
$d_{0}(k) \sin k_{r} = d_{r}(k) \cos k_{r}$ for the components of $k$.
$\Delta = 0$ is obtained if $d_{0}(k)=0$ and $d_{r}(k)=0$ for all $r$.
The latter implies that $k_{r} = 0$ or $k_{r} = \pi$, i.e., the gap closes at the high-symmetry points (HSPs)  in the BZ, which are given by $k_{\rm c} = k_{n_{0}} = (0,...,0,\pi, ..., \pi)$ and by the $D \choose n_{0}$ inequivalent permutations of the components. 
We define $n_{0}$ as the number of vanishing entries $k_{r}=0$.
Then, the first condition $d_{0}(k)=0$ reads:
$0 = m + t \sum_{r=1}^{D} \cos k_{r} = m + t n_{0} - t (D-n_{0}) = m - (D-2n_{0}) t$. 
Using the scaling $t=t^{\ast}/\sqrt{D}$, we find the gap-closure condition $m=\sqrt{D}(1-2n_{0}/D) t^{\ast}$ for $n_{0}=0,...,D$.

The partial orbital-dependent free ($U=0$) density of states is given in terms of the free retarded Green's function as
\be
\rho_{\alpha}(\omega) 
=
- \frac{1}{\pi} \, \mbox{Im} \, G^{(0)}_{\alpha\alpha}(\omega + i0^{+} + \mu^{(0)})
\: ,
\ee
where $\mu^{(0)}$ is the chemical potential of the free system. 
At half-filling, $\mu^{(0)} = 0$, and we have
\be
G^{(0)}_{\alpha\alpha}(\omega)
=
\frac{1}{L} \sum_{k} 
G^{(0)}_{\alpha\alpha}(k,\omega)
\ee
with
\be
G^{(0)}_{\alpha\alpha}(k,\omega)
=
\left[
\frac{1}{\omega - \ff \epsilon(k)}
\right]_{\alpha\alpha} 
=
\left[
\frac{1}{\omega - \sum_{\mu} d_{\mu}(k) \ff \gamma_{D}^{\mu}}
\right]_{\alpha\alpha} \:  .
\ee
Here, $\mu = 0, 1, ... , D$, and $\alpha=1,...,M$ is the orbital index.
The matrix inverse must be computed in the $M=2^{D/2}$-dimensional orbital space.
Exploiting the Clifford-algebra relations again, we get
\be
G^{(0)}_{\alpha\alpha}(k,\omega)
=
\left[
\frac{\omega + \sum_{\mu} d_{\mu}(k) \ff \gamma_{D}^{\mu}}
{\omega^{2} - \sum_{\mu} d_{\mu}(k)^{2}}
\right]_{\alpha\alpha}
\: .
\ee
The irreducible matrix representation of $\ff \gamma^{(r)}$ has vanishing diagonal elements, see Eq.\ (\ref{eq:gamma}), such that there is a contribution from the chiral element $\gamma^{(0)}$ only. 
We have $\gamma^{(0)}_{\alpha\alpha} \equiv z_{\alpha} = +1$ for ``A orbitals'' $\alpha=1,...,M/2$ and 
$\gamma^{(0)}_{\alpha\alpha} \equiv z_{\alpha} = -1$ for ``B orbitals'' $\alpha=(M/2) + 1,...,M$.
This implies that the orbital-resolved free Green's functions $G^{(0)}_{\alpha\alpha}(k,\omega)$ for $\alpha=1,...,M$ can be divided into two classes with representatives $G^{(0)}_{A}(k,\omega)$ and $G^{(0)}_{B}(k,\omega)$.

%%%%%%%%%%%%%%%%%%%%%%%%%%%%%%%%%%%%%%%%%%%%%%%%%%%
\begin{figure}[t]
\includegraphics[width=0.85\columnwidth]{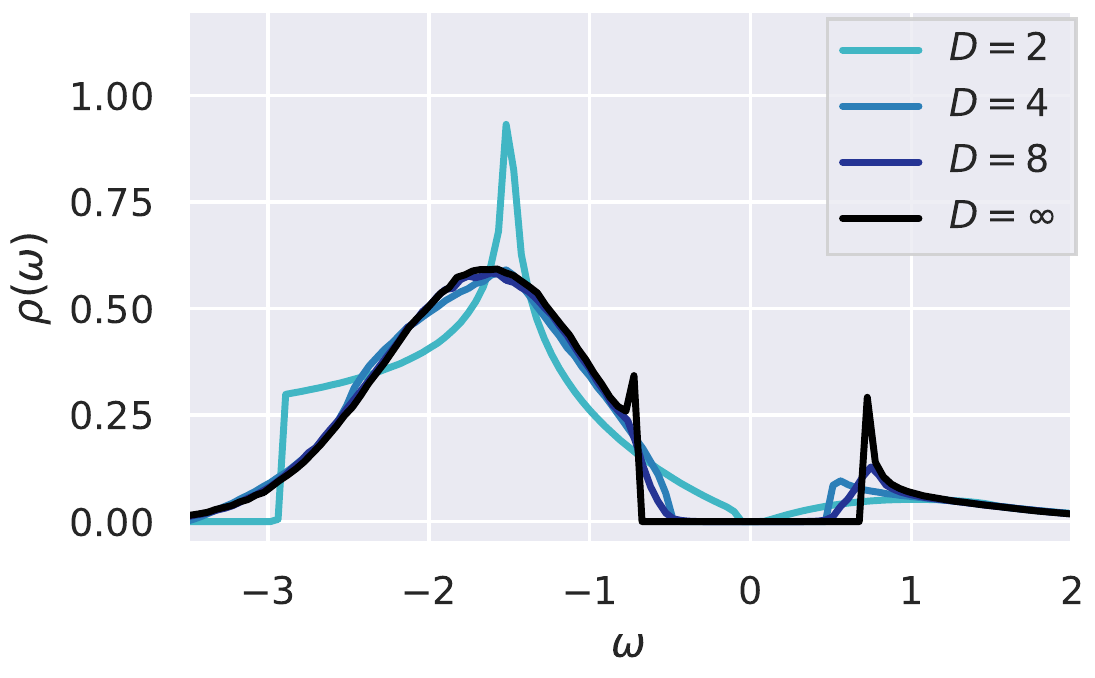}
\caption{
Density of states on the A-orbitals at $m=-1.5$ for various dimensions $D$.
The nearest-neighbor hopping is $t = t^{\ast} / \sqrt{D}$, and $t^{\ast}=1$ sets the energy scale.
}
\label{dosd}
\end{figure}
%%%%%%%%%%%%%%%%%%%%%%%%%%%%%%%%%%%%%%%%%%%%%%%%%%%

%%%%%%%%%%%%%%%%%%%%%%%%%%%%%%%%%%%%%%%%%%%%%%%%%%%
\begin{figure}[t]
\includegraphics[width=0.85\columnwidth]{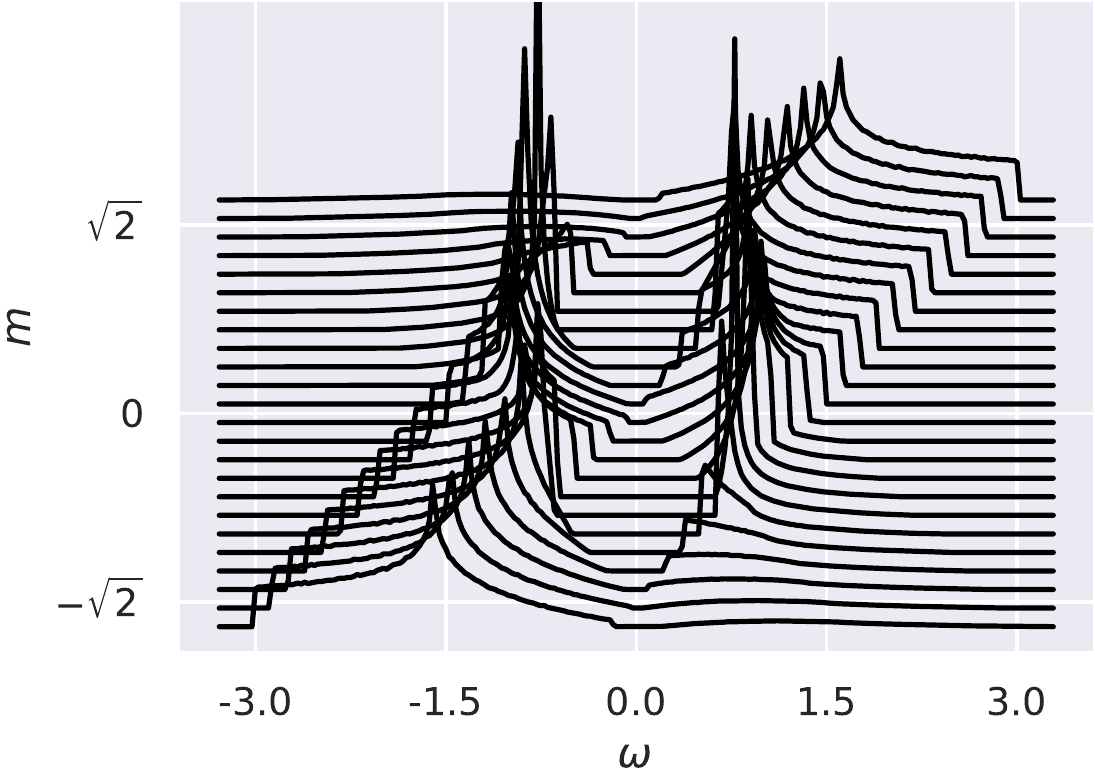}
\caption{
Density of states on the A-orbitals for the $D=2$-dimensional model and various mass parameters $m$. 
}
\label{dosm}
\end{figure}
%%%%%%%%%%%%%%%%%%%%%%%%%%%%%%%%%%%%%%%%%%%%%%%%%%%

Using this result and partial fractional decomposition to get the Lehmann representation of $G^{(0)}_{\alpha\alpha}(k,\omega)$ and finally inserting the result in the expressions above, one finds
\be
\rho_{\alpha}(\omega)
=
\frac{1}{2} \frac{1}{L} \sum_{k} 
\sum_{s=\pm} \left(
1 + s z_{\alpha} \frac{d_{0}(k)}{\epsilon(k)}
\right)
\delta(\omega - s \epsilon(k))
\: , 
\label{eq:dos0}
\ee
where
$\epsilon(k) \equiv \epsilon_{+}(k) = (d_{0}(k)^{2} + \sum_{r} d_{r}(k)^{2})^{1/2}$.
Note that $\frac{1}{M} \sum_{\alpha} \rho_{\alpha}(\omega) 
= 
\frac{1}{L} \sum_{k} 
\frac{1}{2} 
\sum_{s=\pm} \delta(\omega - s \epsilon(k))$. 
Furthermore, the DOS is spin-independent and independent of $\alpha$ for orbitals in the same class A or B. 
Moreover, we have the symmetry $\rho_{A}(-\omega) = \rho_{B}(\omega)$. 
For $m=0$ in particular, $\rho_{\alpha}(-\omega) = \rho_{\alpha}(\omega)$.
Under a sign change $m \to - m$, the DOS transforms as $\rho_{A,B}(\omega) \to \rho_{B,A}(\omega)$.

Fig.\ \ref{dosd} displays the DOS on the $A$ orbitals for a fixed mass parameter $m=-1.5$.
This $m$-value is not critical for any of the finite lattice dimensions considered ($D=2,4,8$), see also Fig.\ \ref{band} (middle).
The DOS at critical mass parameters is discussed in Sec.\ C.
Here, we find a quick overall convergence of the DOS with increasing $D$.

Fig.\ \ref{dosm} gives another example. 
Here, we consider the two-dimensional model, i.e., the model proposed by Qi, Wu and Zhang \cite{QWZ06,AOP16}.
The DOS on the $A$ orbitals is plotted for various mass parameters.
We see the symmetry $\rho_{A}(\omega) \to \rho_{A}(-\omega)$ for $m \to - m$.
Furthermore, the evolution of the gap with $m$ can be read off. 
Gap closures are found at $m=0$ and $m=\pm \sqrt{2}$ (in units of $t^{\ast}=1$).
For $m=0$, the gap closes at $k_{n_{0}=1} = (0,\pi)$ (and at $(\pi, 0)$).
For $m=+1$ and for $m=-1$, the critical $k$-points in the BZ are
$k_{n_{0}=2} = (0,0)$ and $k_{n_{0}=0} = (\pi,\pi)$, respectively.
The $m$-parameter range $0<m<\sqrt{2}$ is characterized by $n_{0}=0$ corresponding to the critical $k$-point for the upper boundary $m=\sqrt{2}$, and the Chern number in that $m$-range is $C_{D=2}(n_{0}=0) = -1$, see also Fig.\ \ref{band} (middle panel) and Eq.\ (\ref{ch}).
For the range $-\sqrt{2} < m<0$, characterized by $n_{0}=1$,  the Chern number is $C_{D=2}(n_{0}=1)=+1$.
The phases for $m < -\sqrt{2}$ and for $\sqrt{2}< m$ are topologically trivial, and the corresponding Chern number vanishes. 

\begin{widetext}

%%%%%%%%%%%%%%%%%%%%%%%%%%%%
\paragraph{\color{blue} 
Section B: DOS in the limit $D\to\infty$.} 
In the limit $D \to \infty$ an analytical expression for the DOS can be given. 
As the $k$-dependence in Eq.\ (\ref{eq:dos0}) is only due to $d_{0}(k)$ and $\sum_{r} d^{2}_{r}(k)$, we can write
\be
\rho_{\alpha}(\omega)
=
\frac{1}{2} 
\sum_{s=\pm} \iint dx dy D(x,y)
\left(
1 + s z_{\alpha} \frac{x}{\sqrt{x^{2}+y}}
\right)
\delta(\omega - s \sqrt{x^{2} + y})
\: 
\label{eq:dos1}
\ee
with 
\be
D(x,y) \equiv \frac{1}{L} \sum_{k} \delta(x-d_{0}(k)) \, \delta(y-\sum_{r} d_{r}^{2}(k)) 
= \frac{1}{(2\pi)^{2}} \iint du dv e^{-iux} e^{-ivy} \Phi(u,v) 
\: .
\ee
In the thermodynamic limit $L\to \infty$ the Fourier transform can be written as:
\be
\Phi(u,v)
= 
e^{ium}\left(
\frac{1}{2\pi} \int_{-\pi}^{\pi} dk \, e^{iut \cos k} e^{ivt^{2} \sin^{2}k}
\right)^{D}
\: . 
\ee
We insert the scaling $t=t^{\ast}/\sqrt{D}$, proceed by straightforwardly expanding the exponentials in powers of $u$ and $v$ and keep terms up to order $1/D$. 
In the limit $D\to \infty$ this yields:
\be
\Phi(u,v) = e^{ium} \, e^{-\frac{1}{4}u^{2} t^{\ast 2}} e^{\frac{1}{2} ivt^{\ast 2}}
\ee
and thus
\be
D(x,y) = \frac{1}{t^{\ast} \sqrt{\pi}} \, e^{-(x-m)^{2}/t^{\ast 2}}
\, \delta(y - \frac{1}{2}t^{\ast 2})
\: .
\label{eq:dxy}
\ee
We see that $D(x,y)$ factorizes for $D\to \infty$.
The computation is a generalization of the one given by M\"uller-Hartmann \cite{MH89b,GKKR96}.
Inserting the result in Eq.\ (\ref{eq:dos1}) we get, after some straightforward algebra: 
\be
\rho_{\alpha}(\omega) 
=
\frac{1}{2} \frac{1}{t^{\ast} \sqrt{\pi}}
\Theta(|\omega| - \frac{1}{\sqrt{2}} t^{\ast})
\, \mbox{sign} \, \omega \sum_{s=\pm}
\left(
\frac{\omega}{\sqrt{\omega^{2} - \frac12 t^{\ast 2}} } + s z_{\alpha}
\right)
\exp\left(
- \frac{\left( 
s \sqrt{\omega^{2} - \frac12 t^{\ast 2}} - m
\right)^{2}}{t^{\ast 2}}
\right)
\: .
\ee
The DOS is has an $m$-independent gap $\Delta=\sqrt{2} t^{\ast}$.

\end{widetext}

%%%%%%%%%%%%%%%%%%%%%%%%%%%%
\paragraph{\color{blue} 
Section C: DOS at a critical $m$.} 

If $m$ is critical, i.e., if the condition for a topological phase transition, $m=\sqrt{D}(1-2n_{0}/D) t^{\ast}$, is satisfied for some $n_{0} \in \{0,...,D\}$, we have $\ff \epsilon(k) = (t^{\ast}/\sqrt{D}) \sum_{r} (k_{r} - k_{n_{0},r}) \ff \gamma^{(r)}_{D}$ close to $\omega=0$ and $k_{n_{0}}$, and the dispersion is given by $d_{0}(k)=0$ and $d_{r}(k)=t(k_{r} - k_{n_{0},r})$, i.e., by a Dirac cone
$\epsilon_{\pm}(k) = \pm (t^{\ast}/\sqrt{D}) [ \sum_{r} (k_{r} - k_{n_{0},r})^{2} ]^{1/2}$.

\begin{figure}[b]
\includegraphics[width=0.85\columnwidth]{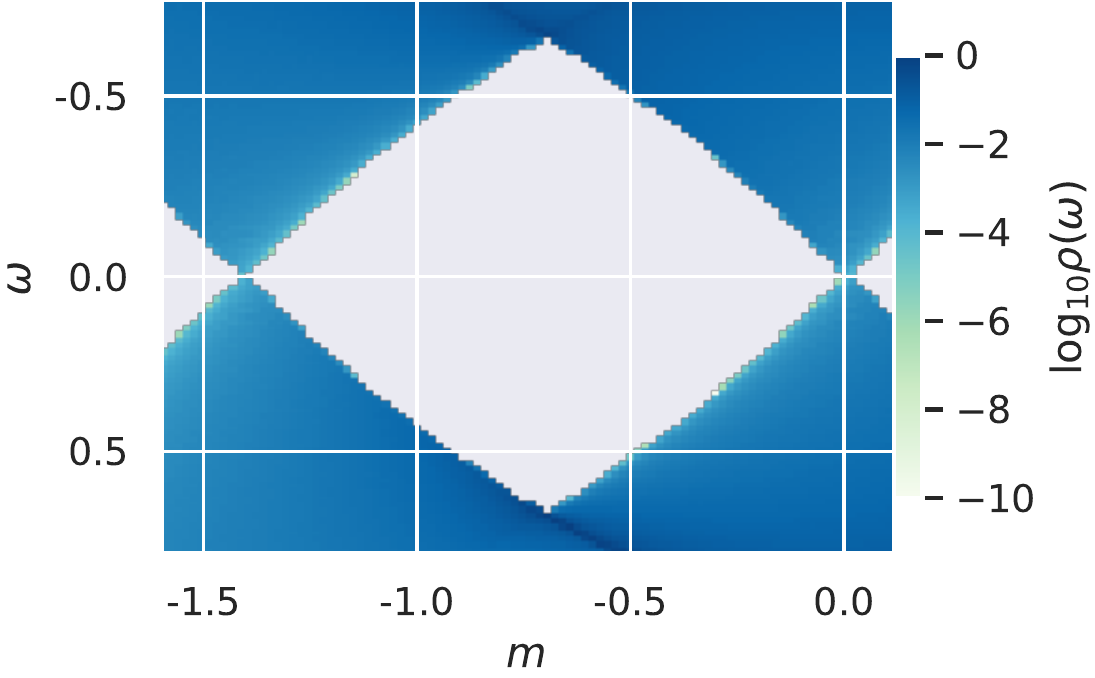}
\caption{
Density of states (color code, note the log-scale) on the A-orbitals at low excitation frequencies as a function of the mass parameter $m$. Calculation for the $D=2$-dimensional model.
}
\label{dosg}
\end{figure}

Fig.\ \ref{dosg} provides an overview for the $D=2$ model and for the gap closures at $m=-\sqrt{2}$ and $m=0$ (in units of $t^{\ast}=1$).
Directly at the critical $m$, and at low frequencies, the DOS is linear $\rho_{A}(\omega) \propto |\omega|$. 
The figure shows that the $m$-dependence of the gap is linear as well, $\Delta \propto (m-m_{c})$, if $m$ is sufficiently close to a critical value $m_{c}$.

%%%%%%%%%%%%%%%%%%%%%%%%%%%%%%%%%%%%%%%%%%%%%%%%%%%
%%%%%%%%%%%%%%%%%%%%%%%%%%%%%%%%%%%%%%%%%%%%%%%%%%%

We proceed with an analytical calculation for arbitrary $D$.
The low-frequency DOS for the $D$-dimensional model is still given by Eq.\ (\ref{eq:dos1}) but with 
\be
D(x,y) \equiv \delta(x) \, \frac{1}{L} \, {\sum_{k}}' \delta \left (y - \sum_{r} d_{r}^{2}(k)  \right)
\: ,
\ee
where $\sum'_{k}$ indicates summation over wave vectors differences $k$ with respect to a high-symmetry point $k_{\rm c} = k_{n_{0}} = (0,...,0,\pi, ..., \pi)$ within a sphere $|k| \le \Lambda$ defined by a cutoff $\Lambda$.
This implies
\be
\rho_{\alpha}(\omega)
=
{D \choose n_{0}}
\frac{1}{2} 
\sum_{s=\pm} 
\frac{1}{L} \, {\sum_{k}}'
\delta\left(\omega - s \frac{t^{\ast}}{\sqrt{D}} \sqrt{\sum_{r=1}^{D} k_{r}^{2} } \right)
\:
\label{eq:doslin2}
\ee
in the linear low-frequency regime. 
The combinatorial prefactor accounts for the fact that the gap closes simultaneously at all $D \choose n_{0}$ wave vectors produced by the permutations of the components of $k_{\rm c} = k_{n_{0}} = (0,...,0,\pi, ..., \pi)$.
In the thermodynamic limit, and at sufficiently low frequencies $\omega$, 
\ba
\rho_{\alpha}(\omega)
&=&
{D \choose n_{0}}
\frac{1}{2} 
\sum_{s=\pm} 
\frac{S_{D-1}}{(2\pi)^{D}}
\int_{0}^{\Lambda} d \kappa \, \kappa^{D-1}
\delta\left(\omega -  \frac{s t^{\ast}}{\sqrt{D}} \kappa \right)
\nonumber \\
&=&
\frac{|\omega|^{D-1}}{t^{\ast D}} \frac{1}{2^{D}} \frac{D!}{n_{0}!(D-n_{0})!} \frac{D^{D/2}}{(D/2-1)!} \frac{1}{\pi^{D/2}}
\: ,
\label{eq:doslin1}
\ea
with $\kappa=|k|$, and with the surface area $S_{D-1} = 2 \pi^{\nicefrac D2}/(\nicefrac D2-1)! = 2 \pi^{\nicefrac D2}/\Gamma(\nicefrac D2)$ of the $D-1$-dimensional unit sphere $S^{D-1}$. 
This implies
\be
\rho_{\alpha}(\omega)
=
c(D,n_{0}) |\omega|^{D-1} / t^{\ast D}
\ee
at low frequencies with a coefficient $c(D,n_{0})$ which, for any $n_{0}$, tends to zero exponentially fast as $D\to \infty$.
\\

%%%%%%%%%%%%%%%%%%%%%%%%%%%%
\paragraph{\color{blue} 
Section D: Diagonal elements of the spectral function.} 
With the help of the self-energy, the interacting Green's function generally reads
\be
  \ff G(k,\omega) = \frac{1}{\omega + \mu - \ff \epsilon(k) - \ff \Sigma(k,\omega)} \: .
\label{eq:green}
\ee
Here, we have included a chemical-potential term in the Hamiltonian via the replacement $\ff \epsilon(k) \mapsto \ff \epsilon(k) - \mu \ff 1$. 

As the interaction term preserves the symmetries at half-filling, we must have $A_{A/B}(\omega) = A_{B/A}(-\omega)$ for the interacting local spectral function, $A_{\alpha}(\omega)=-(1/\pi L)\mbox{Im} \sum_{k} G_{\alpha\alpha\sigma\sigma}(k,\omega+i0^{+})$.
This implies that the total ($\alpha$-summed) local spectral density $A(\omega)$ is symmetric. 
Hence, half-filling is obtained with a chemical potential which yields a vanishing first moment of $A(\omega)$.
The latter is given by $M_{\alpha}^{(1)} = \frac{1}{M} \sum_{\alpha } (m \gamma^{(0)}_{\alpha\alpha} + U \langle n_{\alpha} \rangle - \mu)$, i.e., we must choose $\mu = U/2$ since the orbital occupations must be symmetric as well: $\langle n_{A} \rangle + \langle n_{B} \rangle = 1$.

Within the DMFT, the self-energy is site-diagonal, i.e., $k$-independent. 
Furthermore, as the Hubbard-interaction term is an intra-orbital interaction only, it is diagonal in orbital space, 
\be
  \Sigma_{\alpha\beta}(k,\omega) = \Sigma_{\alpha\beta}(\omega) 
  = \delta_{\alpha\beta} \Sigma_{\alpha}(\omega) 
  \: .
\ee
Analogous to the discussion of the density of states above, the orbital-dependent diagonal elements $\Sigma_{\alpha}(\omega)$ can be divided into two classes $A$ and $B$.
With the definition
\be
\Sigma_{\pm}(\omega) 
= 
\frac{1}{2} \left( 
\Sigma_{A} (\omega) \pm \Sigma_{B}(\omega)
\right) \:,
\ee
we have the following decomposition:
\be
\ff \Sigma(\omega) 
=
\Sigma_{+}(\omega) \ff 1
+
\Sigma_{-}(\omega) \ff \gamma_{D}^{(0)}
\: .
\label{sigmapm}
\ee
\begin{widetext}
Inserting this into Eq.\ (\ref{eq:green}), we can treat the matrix inversion analogously to the non-interacting case to get the local Green's function $G_{\alpha}(\omega) \equiv \frac{1}{L} \sum_{k} G_{\alpha\alpha}(k,\omega)$ of orbital $\alpha$ in the form
\be
G_{\alpha}(\omega)
=
\frac{1}{2L} \sum_{k, s=\pm} 
\frac{
1+s z_{\alpha}
\frac{
d_{0}(k)+ \Sigma_{-}(\omega)
}
{
\sqrt{
\sum_{r} d^{2}_{r}(k) + \left(d_{0}(k) + \Sigma_{-}(\omega) \right)^{2}
}
}
}{
\omega + \mu - \Sigma_{+}(\omega) - s \sqrt{
\sum_{r} d^{2}_{r}(k) + \left(d_{0}(k) + \Sigma_{-}(\omega) \right)^{2}
}
}
\: . 
\ee
In the limit $D\to \infty$ and using Eq.\ (\ref{eq:dxy}), this can be written as:
\be
G_{\alpha}(\omega)
=
\frac{1}{2} \sum_{s=\pm} \frac{1}{t^{\ast} \sqrt{\pi}}
\int dx
\frac{
1+s z_{\alpha}
\frac{
x + \Sigma_{-}(\omega)
}
{
\sqrt{
\frac12 t^{\ast 2} + \left( x + \Sigma_{-}(\omega) \right)^{2}
}
}
}{
\omega + \mu - \Sigma_{+}(\omega) - s \sqrt{
\frac12 t^{\ast 2} + \left(x + \Sigma_{-}(\omega) \right)^{2}
}
}
\,
\exp{\left( 
- (x-m)^{2} / t^{\ast 2}
\right)}
\: . 
\label{gint}
\ee

\end{widetext}

%%%%%%%%%%%%%%%%%%%%%%%%%%%%
\paragraph{\color{blue} 
Section E: Two-site DMFT.} 

A simplified variant of dynamical mean-field theory is helpful for an efficient computation of the whole $m$-$U$ phase diagram. 
The general theory is explained and discussed in Ref.\ \cite{Pot01a}. 
Here, we just present the basic operational steps necessary for the concrete numerical computations and also a slight generalization of the approach to the multi-orbital case with intra-orbital Hubbard interaction and the staggered orbital field introduced by the mass term.

Within two-site DMFT, the interacting lattice model is self-consistently mapped onto a single-impurity Anderson model (SIAM) for A orbitals and to another one for B orbitals. 
Each $\alpha$-SIAM ($\alpha = \mbox{A,B}$) consists of two sites only: a spin-degenerate, correlated impurity site with one-particle energy $\varepsilon_{d,\alpha}= m \gamma^{(0)}_{\alpha\alpha} = \pm m$ and with the Hubbard interaction present, and a spin-degenerate, noninteracting bath site with one-particle energy $\varepsilon_{c,\alpha}$. 
The sites are coupled by a spin-indepenent hybridization term of strength $V_{\alpha}$.

Due to the small Hilbert space, the two-site SIAM can be solved easily for the single-particle impurity Green's function $G_{\rm imp, \alpha}(\omega)$, from which we obtain the ground-state occupation of the impurity site $n_{\rm imp, \alpha}$ and the self-energy $\Sigma_{\alpha}(\omega)$, which is local and nonzero at the impurity site only and identified with the local lattice self-energy of orbital $\alpha$.
Furthermore, we compute the renormalization factor
$z_{\alpha} = 1/(1 - \partial \Sigma_{\alpha}(\omega=0) / \partial\omega)$.
For the system considered here, we have the symmetry relation
$\Sigma_{-}(\omega) = \Sigma_{-}(-\omega)$ for $\Sigma_-(\omega) \equiv (\Sigma_{A}(\omega) -\Sigma_{B}(\omega))/2$.  
With Eq.\ (\ref{sigmapm}) this implies that $z_{\alpha}=z=\mbox{const}=
1/(1 - \partial \Sigma_{+}(\omega=0) / \partial\omega)$.

With the self-energy at hand, Eq.\ (\ref{gint}) provides us with the local element of the interacting lattice Green's function $G_{\alpha}(\omega)$.
The original DMFT self-consistency requires $G_{\alpha}(\omega) \stackrel{!}{=} G_{\rm imp,\alpha}(\omega)$. 
This, however, will be relaxed as there are only two (spin-independent) parameters to be fixed. 
$V_{\alpha}$ and $\varepsilon_{c,\alpha}$ are self-consistently determined from the following two two-site DMFT self-consistency conditions (for each $\alpha = \mbox{A,B}$): 
\be
  n_{\rm imp,\alpha} \stackrel{!}{=} n_{\rm \alpha} \: , 
\label{sc1}
\ee
where $n_{\alpha}$ is the ground-state occupation of orbital $\alpha$ in the lattice model that is obtained from $G_{\alpha}(\omega)$,
and 
\be
  V_{\alpha}^{2} \stackrel{!}{=} z_{\alpha} M_{\alpha}^{(2)} - z_{\alpha} {M_{\alpha}^{(1)}}^{2} = z t^{\ast 2} \; ,
\label{sc2} 
\ee
i.e., $V_{\alpha}^{2}$ is fixed by the second, centered moment of the noninteracting DOS and the renormalization factor $z$.

Condition (\ref{sc1}) ensures that all coefficients in the high-frequency expansion of $G_{\alpha}(\omega)$ are exact up to and including terms or order $1/\omega^{3}$.
Therewith, the norm, the center of gravity, and the variance of the interacting local spectral density $- \mbox{Im}G_{\alpha}(\omega+i0^{+})/\pi$ are reproduced correctly.
This is important to assure the correct positions and weights of the Hubbard bands at high frequencies.

\begin{widetext}
Condition (\ref{sc2}) is obtained as follows: 
We use
$\ff \Sigma(\omega) = \ff a + (1 - z^{-1}) \omega +\ca O(\omega^{2})$ in Dyson's equation to get the ``coherent'' Green's function
\be
{\ff G}^{(\rm low)}(k,\omega) 
\equiv
\frac{1}{z^{-1}\omega + \mu - \ff a - \ff \epsilon(k)}
=
\frac{z}{\omega + z (\mu - \ff a - \ff \epsilon(k) )}
\: , 
\ee
where $\ff a \equiv \ff \Sigma(0)$ is diagonal. 
${\ff G}^{(\rm low)}(k,\omega)$ describes the low-energy excitations. 
For a metallic phase, the $z$-factor is the weight of the quasi-particle resonance at the Fermi energy \cite{Pot01a}. 
Here, for insulating phases, the $z$-factor describes the gap renomalization, see Ref.\ \cite{SKWK09}.

The first three coefficients of the high-frequency expansion of ${\ff G}^{(\rm low)}(k,\omega)$ fix the norm, the center of gravity, and the variance of the gapped low-frequency spectrum.
We have:
\be
\frac{1}{L} \sum_{k} {\ff G}^{(\rm low)}(k,\omega)
=
\frac{z}{\omega}
+
\frac{z^{2} (\ff M^{(1)} + \ff a - \mu)}{\omega^{2}}
+
\frac{
z^{3}(
\ff M^{(2)}
+
(\ff M^{(1)} + \ff a - \mu)^2 - {\ff M^{(1)}}^{2}
)}{\omega^{3}}
+ \ca O(\omega^{4})
\: ,
\label{low1}
\ee
with $M^{(1)}_{\alpha\alpha'} = m \gamma^{(0)}_{\alpha\alpha'} = \pm m \delta_{\alpha\alpha'}$ and
$M^{(2)}_{\alpha\alpha'} = (t^{\ast 2} + m^2) \delta_{\alpha\alpha'}$.
DMFT self-consistency requires that the local elements 
${G}^{(\rm low)}_{\alpha\alpha}(\omega) = L^{-1} \sum_{k} {G}^{(\rm low)}_{\alpha\alpha}(k,\omega)$
equal the low-energy coherent impurity Green's functions $G^{(\rm imp, low)}_{\alpha}(\omega)$.
Performing the high-frequency expansion of $G^{(\rm imp, low)}_{\alpha}(\omega)$ for the $\alpha$-th SIAM yields \cite{Pot01a}:
\be
G^{(\rm imp, low)}_{\alpha}(\omega)
= 
\frac{z_{\alpha}}{\omega}
+
\frac{z_{\alpha}^{2} (\varepsilon_{d, \alpha} + a_{\alpha} - \mu)}{\omega^{2}}
+
\frac{
z_{\alpha}^{2} V_{\alpha}^{2}
+
z_{\alpha}^{3} (\varepsilon_{d, \alpha} + a_{\alpha} - \mu)^{2}}
{\omega^{3}}
+ \ca O(\omega^{4})
\: ,
\label{low2}
\ee
Comparing Eqs.\ (\ref{low1}) with (\ref{low2}) and noting that $z_{A}=z_{B}$ and $\varepsilon_{d, \alpha}= M_{\alpha\alpha}^{(1)}=M_{\alpha}^{(1)}$, yields Eq.\ (\ref{sc2}).

\end{widetext}

%%%%%%%%%%%%%%%%%%%%%%%%%%%%
\paragraph{\color{blue} 
Section F: Further numerical results.} 
Here, we present some further results obtained with the two-site DMFT for the interacting system.

%%%%%%%%%%%%%%%%%%%%%%%%%%%%%%%%%%%%%%%%%%%%%%%%%%%
\begin{figure}[b]
\includegraphics[width=0.95\columnwidth]{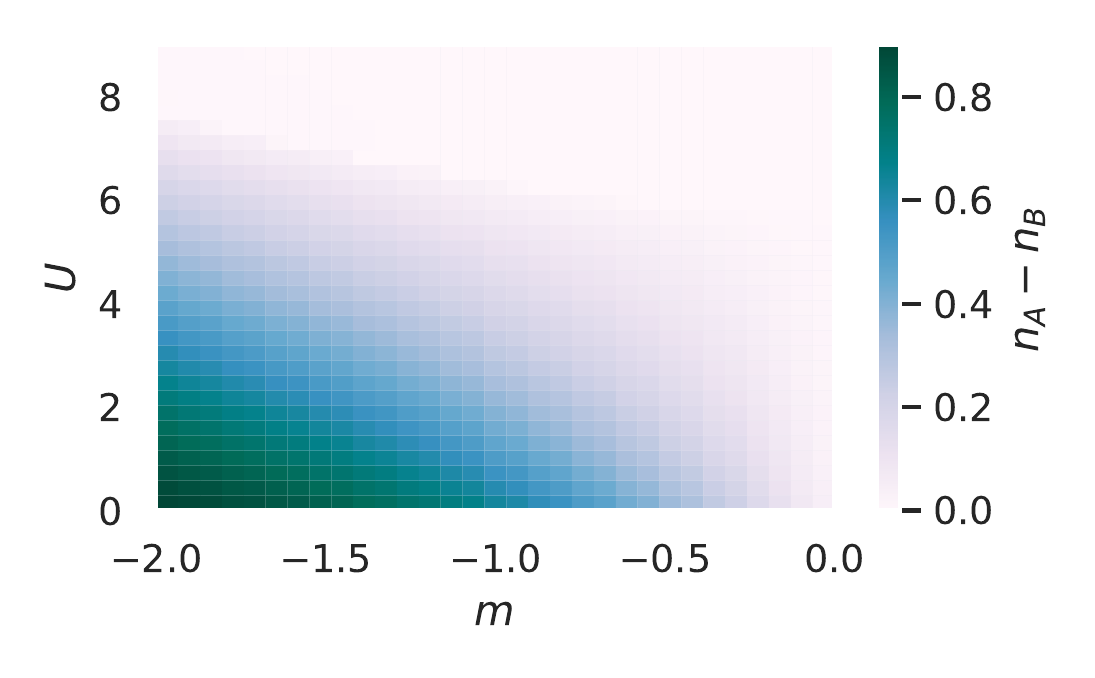}
\caption{
$m$-$U$-dependence of the orbital polarization $n_{A}-n_{B}$ in the $D\to \infty$ limit.
}
\label{polu}
\end{figure}
%%%%%%%%%%%%%%%%%%%%%%%%%%%%%%%%%%%%%%%%%%%%%%%%%%%

Fig.\ \ref{polu} displays the orbital polarization $n_{A}-n_{B}$ as a function of $m$ and $U$ in the $D\to\infty$ limit. 
At $m=0$ and for all $U$ both types of orbitals are equally occupied. 
Though this is hardly visible in the figure, the raw data shows, that the same holds for the entire Mott-insulating phase. 
On the contrary, the system is most easily polarizable at $U=0$. 

%%%%%%%%%%%%%%%%%%%%%%%%%%%%%%%%%%%%%%%%%%%%%%%%%%%
\begin{figure}[b]
\includegraphics[width=0.95\columnwidth]{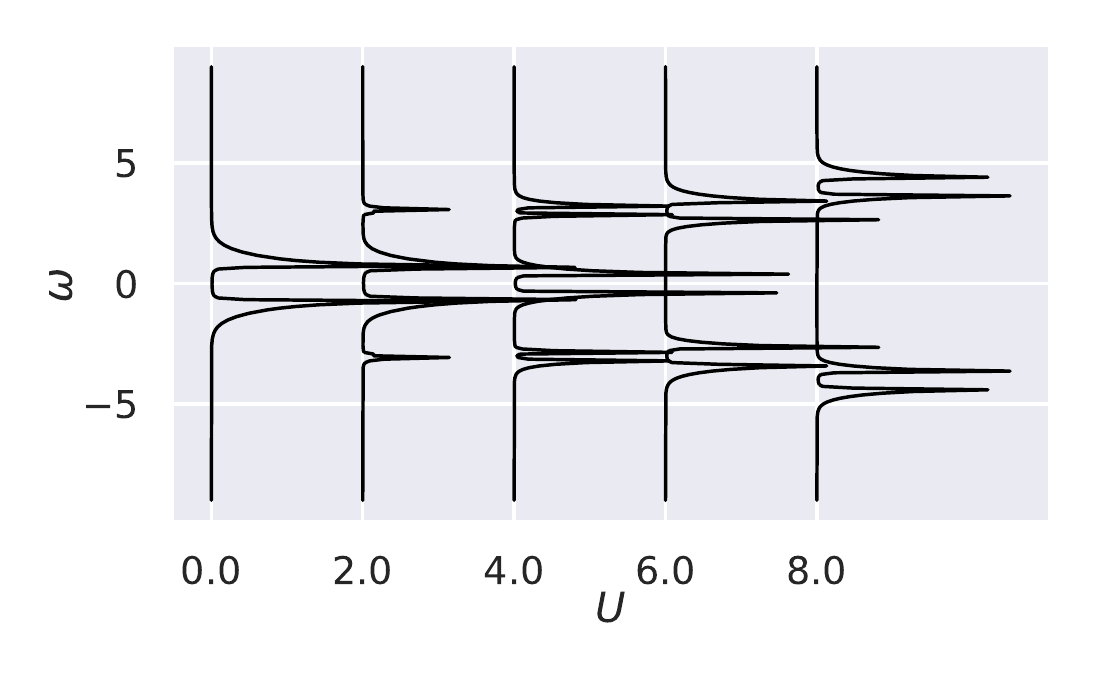}
\includegraphics[width=0.95\columnwidth]{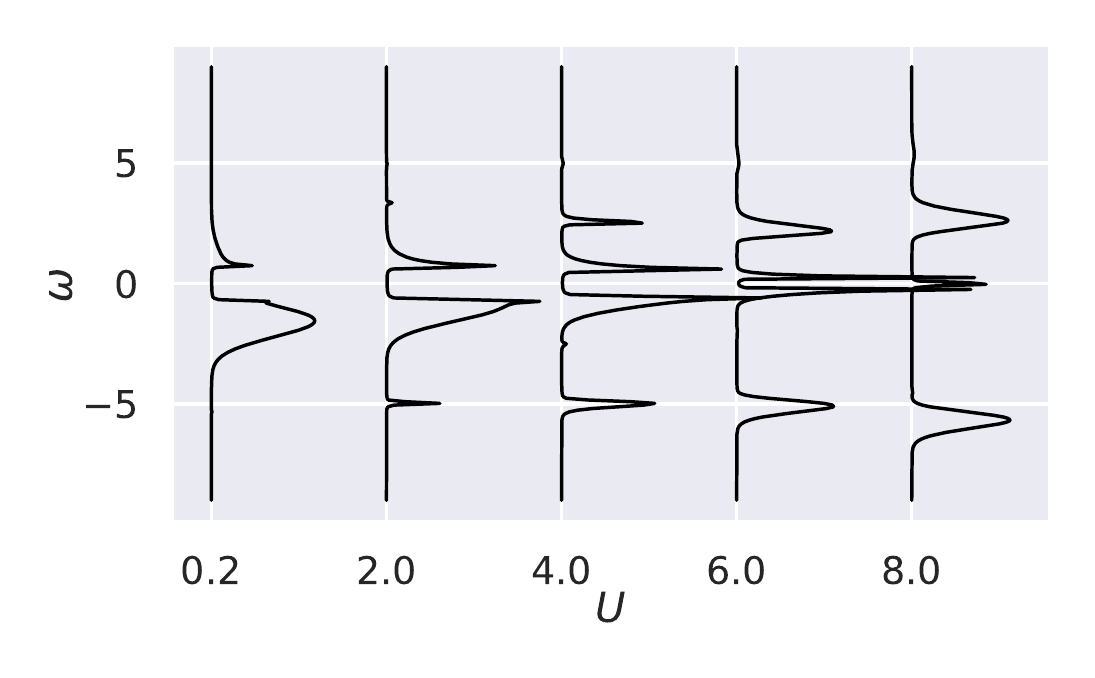}
\caption{
Local spectral function $A^{\rm (loc)}_{\alpha\alpha}(\omega)$ on $\alpha=A$-orbitals for $m=0$ (top) and $m=-1.5$ (bottom) for various $U$ in the $D\to \infty$ limit.
}
\label{alocal}
\end{figure}
%%%%%%%%%%%%%%%%%%%%%%%%%%%%%%%%%%%%%%%%%%%%%%%%%%%

Fig.\ \ref{alocal} shows the local spectral function on the A-orbitals for various $U$ in the symmetric case at $m=0$ and for $m=-1.5$. 
At $m=0$, the spectral function is symmetric in $\omega$. 
For $U=0$, the gap is $\Delta=\sqrt{2}t^{\ast}$. 
A slight Lorentzian broadening of the spectrum is artificial and caused a finite imaginary part $\varepsilon=0.01$, introduced via $\omega \mapsto \omega + i \varepsilon$ in the calculation of
$A^{\rm (loc)}_{\alpha\alpha}(\omega) = - (1/\pi) \mbox{Im} \, G^{\rm (loc)}_{\alpha\alpha}(\omega+i\varepsilon)$ from the retarded local Green's function.
With increasing $U$ the gap shrinks. 
This is related to the renormalization factor $z$ which decreases upon approaching the transition to the Mott insulator at $U=U_{\rm c}=6t^{\ast}$.
Close to the transition $U \to U_{\rm c}$, the gap $\Delta(U)\to 0$. 
At the same time the spectral weight of the low-energy peaks in the spectral function vanish, and for $U>U_{\rm c}$ a large Mott-Hubbard gap is present.
This increases with increasing $U$, again related to an increase of $z$ with $U$ for $U> U_{\rm c}$.

For $m=-1.5$, the spectral function is asymmetric in $\omega$, reflecting the orbital polarization of the system. 
Apart from that, the evolution of 
$A^{\rm (loc)}_{\alpha\alpha}(\omega)$ with $U$ is qualitatively the same.
The spectrum for $U=8$ is actually still gapped. 
Due to the finite broadening $\varepsilon$, however, a single artificial peak centered around $\omega=0$ is visible in the figure.
Note that with increasing $U$ there is a strong spectral-weight transfer taking place, such that for $U> U_{\rm c}$ the orbital polarization vanishes.

%%%%%%%%%%%%%%%%%%%%%%%%%%%%%%%%%%%%%%%%%%%%%%%%%%%
\begin{figure}[t]
\includegraphics[width=0.7\columnwidth]{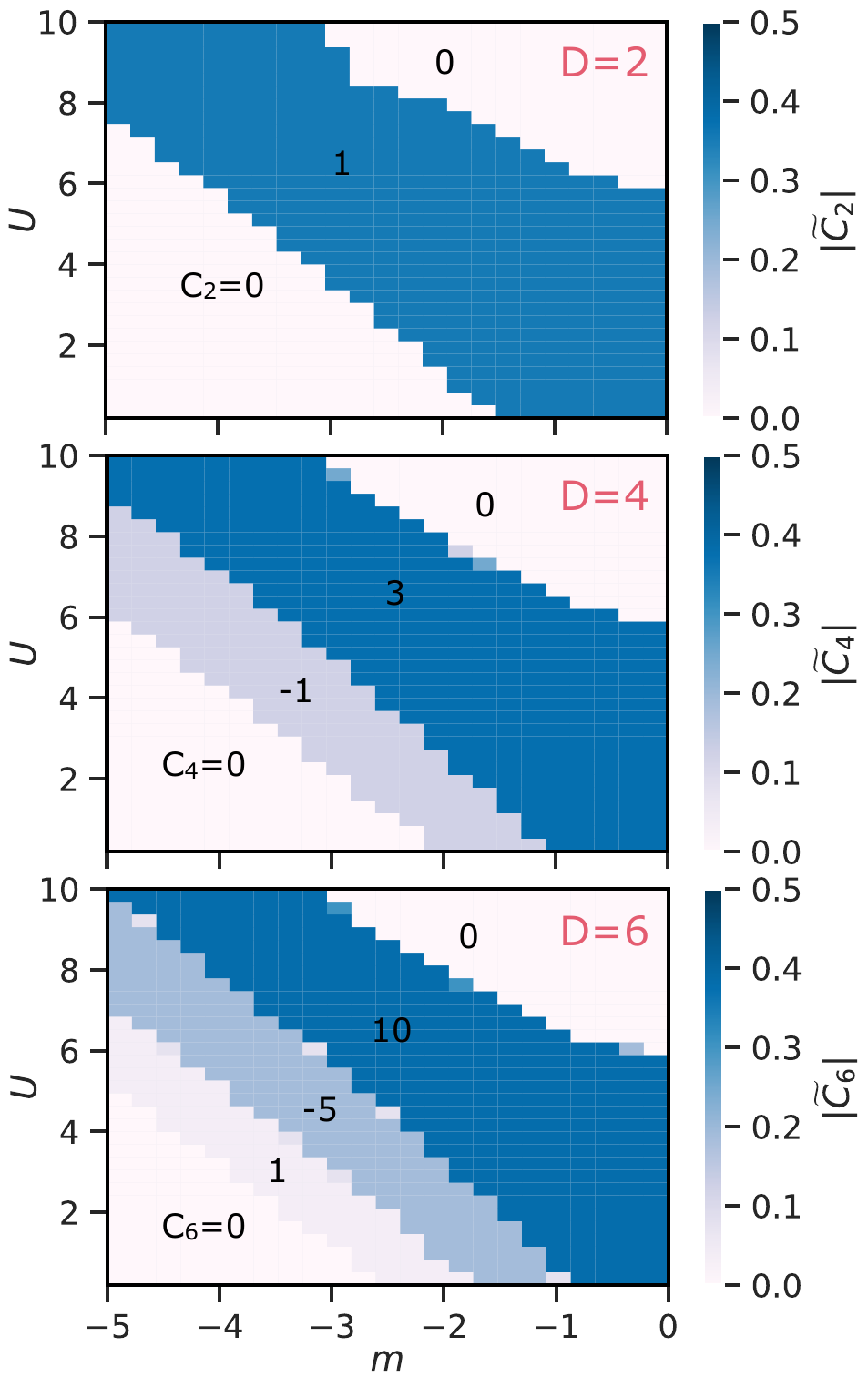}
\caption{
$m$-$U$ phase diagrams for $D=2,4,6$ computed for $t=t^{\ast}/\sqrt{D}$ with $t^{\ast}=1$. 
The color codes the modulus of the normalized Chern number $\widetilde{C}_{D} = C_{D} / 2^{D}$.
The values of the $D$-th Chern numbers for the different phases are indicated in the plot.
}
\label{pd246}
\end{figure}
%%%%%%%%%%%%%%%%%%%%%%%%%%%%%%%%%%%%%%%%%%%%%%%%%%%

Fig.\ \ref{pd246} displays $m$-$U$ phase diagrams for finite dimensions $D=2, 4, 6$ (top to bottom) to be compared with the phase diagram, Fig.\ \ref{pd}, which is discussed in the main text. 
To make the results for different $D$ comparable, we have computed all phase diagrams with the scaled hopping parameter $t=t^{\ast}/\sqrt{D}$ where $t^{\ast}=1$ and have color coded the Chern density $c(m,U)$, which is nonnegative and normalized as discussed in the main text, rather than the Chern number, Eq.\ (\ref{ch}). 
The color coding is the same in all figures, including Fig.\ \ref{pd}.

The Chern number is given additionally and labels the different topological phases in the figure.
Note the alternating sign and the monotonic increase of the Chern number along any straight path from the band to the Mott insulator.

The convergence of $c(m,U)$, opposed to the Chern number, for each point in the entire phase diagram with increasing $D$ to the respective points in the $D=\infty$ phase diagram (Fig.\ \ref{pd}) is highly plausible. 
With increasing $D$, the parameter regions of constant $c(m,U)$ shrink in size. 
This is balanced by an ever-increasing number of phases, such that for $D\to \infty$ a continuum of topologically different phases is obtained, each of which covers a one-dimensional manifold in the $m$-$U$ plane, which is defined by $c(m,U)=\mbox{const}$.

\end{document}